# ON THE HYPERELASTIC BEHAVIOR OF THE BOAR DIAPHRAGMATIC TENDON MEMBRANE BY INFLATION TESTS AND MODELING

Rania Abdel Rahman El Anwar[1,2*], J. Vasquez-Villegas[1], S. Le Floc'h[1], P. Royer[1], N. Bahlouli[3], C. Wagner-Kocher[1]

(1) University of Montpellier, LMGC, CNRS, Montpellier, France
(2) Laboratoire Interdisciplinaire de l'Université Française d'Égypte (UFEID) - Université française d'Égypte, Cairo 11837, Egypt
(3) University of Strasbourg, ICube, CNRS, Strasbourg, France

*Corresponding author: R.A.ElAnwar
Email address: rania.elanwar@ufe.edu.eg

**Abstract**

**Background**: Despite the large variety of materials used to repair congenital diaphragmatic hernia (CDH), none has proven ideal due to complications and risk of recurrence. Understanding the mechanical behavior of the diaphragm's central tendon is essential for developing biomimetic prostheses. **Objective**: This study aims to characterize the hyperelastic behavior of the porcine diaphragmatic tendon under biaxial loading conditions. **Methods**: Biaxial hyperelastic response of the porcine diaphragmatic central tendon was characterized using bulge inflation tests combined with full-field stereo digital image correlation (3D-DIC). Principal stretches were extracted from the reconstructed three-dimensional geometry using a spherical-cap approximation, with corrections for clamping-induced pre-deformation. Several incompressible isotropic hyperelastic models (Neo-Hookean, Mooney–Rivlin, Yeoh and Fung) were first evaluated as phenomenological baselines. To account for the anisotropic mechanical signature of the tendon, a transversely isotropic hyperelastic formulation of the Humphrey-Yin type was implemented. This model represents an effective anisotropic response associated with the lamellar and collagen-rich structure of the tissue through its thickness, while assuming isotropy within the membrane plane. **Results**: The diaphragm tissue exhibited an exponential mechanical response. Neo-Hookean and Mooney-Rivlin models failed to capture the observed behavior, while the Yeoh model slightly overestimated nonlinearity. The Fung model provided the closest fit to the nonlinear pressure–stretch response, yet failed to reproduce the directional differences observed between meridional and circumferential stretches. The Humphrey-Yin model provided a markedly improved description of the experimental data over the full inflation range. Parameter identification revealed that the transversely isotropic contribution dominated the strain-energy response, highlighting the limitations of isotropic constitutive laws for modeling diaphragmatic tendon mechanics, even under macroscopically axisymmetric inflation. Despite specimen-to-specimen variability, no significant differences were observed between left and right diaphragm samples. **Conclusion**: Overall, this work demonstrates that effective anisotropic hyperelastic formulations are required to describe the biaxial mechanical behavior of the diaphragmatic central tendon under inflation loading. The proposed experimental–numerical framework provides a robust basis for biomechanical modeling and constitutes a first step toward the development of biomimetic prosthetic materials for diaphragmatic repair.

## 1. Introduction

The diaphragm is a dome-shaped musculotendinous structure that separates the thoracic and abdominal cavities by closing the inferior thoracic aperture. This delimitation is vital for the proper functioning and development of various organs during pregnancy. The diaphragm acts as the main respiratory muscle. It consists of a peripheral part, consisting of muscular fibers, and a central part, which is predominantly tendon, Fig. (1-a). The tendon forms the upper part of the diaphragm and is in direct contact with the lungs. It is positioned closer to the front of the thorax than the back and has a distinctive trefoil shape with an average surface area of 143 $cm^2$ in humans. The muscle fibers can be described as radiating from the central tendon and merging toward their outer attachments while following the curvature of the muscle surface in its resting configuration [1]. The central tendon is composed of multiple planes of fibers that are arranged in a way that provides strength and rigidity [2]. Moreover, it is relatively inextensible, thus restricting the ability of the diaphragm to change curvature in response to trans diaphragmatic pressure [3, 4]. The diaphragm can be considered as a thin membrane capable of carrying significant in-plane stresses but is not designed to bear bending moments or shear forces out of the plane of the membrane [2].

The diaphragm, being a thin membrane, is susceptible to ruptures due to thoracic or abdominal trauma. According to Favre et al. [5], two mechanisms may explain diaphragmatic ruptures: an increase in abdominal pressure due to frontal impact and deformation of the lower thoracic aperture during lateral impact. Congenital malformations may also interrupt the continuity of the diaphragm tissues, notably congenital diaphragmatic hernia (CDH), which results from incomplete muscle closure during embryonic development [6]. The hernia creates a passage between the thoracic and abdominal cavities, allowing abdominal organs to enter the thorax and impair pulmonary development. In severe cases, the newborn may be left with insufficient pulmonary tissue to survive [7,8]. Surgical repair involves closing the defect either through suturing or prosthetic implantation [9], yet no prosthetic material has proven fully satisfactory due to recurrence and complications [10,11]. Tissue-engineered constructs have been explored, with tendinous implants showing better long-term integration than muscular implants [12,13], which motivates focusing on the mechanical characterization of the central tendon for biomimetic prosthesis development.

Previous studies have indicated that the diaphragm, as a soft biological tissue, is a fibrous anisotropic membrane with collagen fibers aligned in different orientations [14–17]. Therefore, to gain a more precise understanding of its mechanical behavior and facilitate the design of diaphragmatic prostheses that align with its actual behavior, it is essential to conduct multi-axial tests. Although the bulge test has been recognized as a powerful tool to study multiaxial stress and strain fields in biological tissues [18,19] and nonlinear membranes [20–23], no previous work has applied this method to the diaphragm. Instead, uniaxial tensile tests have been performed on human and animal diaphragms, revealing mechanical dependencies on age [24], strain rate [25], and a nonlinear, inelastic, and anisotropic behavior of the central tendon [15]. For this reason, hyperelastic constitutive modeling is required. When combined with Three-dimensional Digital Image Correlation (3D-DIC), the bulge test enables capturing full-field deformation and extracting constitutive parameters from biaxial loading [7,26,27]. Although the macroscopic inflation response appears nearly isotropic at the pole, this behavior should be regarded as a phenomenological approximation rather than a physically representative constitutive description.

Being a non-contact optical method, DIC is ideal for investigating the mechanical properties of soft biological tissues. It provides full-field displacement data, provided that a high-quality stochastic speckle pattern is applied [28-30]. While 2D-DIC uncertainties have been extensively reported, fewer studies address the specific uncertainty sources of 3D-DIC. Reu [31] and Kelleher et al. [32] emphasized contributions from calibration, pattern quality, image noise, and hidden correlation parameters, while Ke et al. [33] studied the influence of stereo angle. Typical uncertainty values in literature range from 0.01 to 0.02 pixels [34,35]. Because constitutive identification from bulge tests is highly sensitive to measurement accuracy, quantifying experimental uncertainty is essential. In the present study, potential error sources were systematically evaluated using Fourier-based pattern quality assessment, controlled-lighting verification, and inter-sample consistency analyses across all twelve specimens. This ensures that the extracted mechanical parameters reflect true material behavior rather than artefacts of the measurement process.

The present work aims to characterize the biaxial deformation behavior of the porcine diaphragmatic central tendon using bulge inflation tests coupled with full-field 3D-DIC measurements. Experimental stretch data are used to identify the parameters of several isotropic hyperelastic models and to evaluate the transversely isotropic formulation, thereby quantifying the role of collagen-fiber reinforcement. The resulting constitutive description is intended to support future biomechanical modeling efforts and guide the design of biomimetic prosthetic materials for diaphragmatic repair.

## 2. Materials and Methods

### *2.1 Samples Preparation:*

One fresh diaphragm was excised from 65 kg wild boar hunted at Aniane (South of France), Fig. (1-a). The diaphragm was immediately divided into right and left halves, Fig. (1-b), and both halves were frozen at -50°C for preservation until testing. Unlike muscular tissue, freezing is known to have almost negligible effects on the mechanical properties of tendons, since they contain only a few cells, provided that repeated freezing–thawing cycles are avoided [36 – 39]. Prior to testing, one-half of the diaphragm was thawed at a time in a saline solution of sodium chloride with a concentration of 0.9 g/L. Circular samples with a diameter of 30 mm were then cut from the central tendon using a stamp and a scalpel, Fig. (1-b). The thickness of each sample was measured at multiple points using optical coherence tomography (OCT), and the average central thickness was evaluated (Table 1). Thickness variations were not explicitly included in the membrane equilibrium equation and constitute a limitation of the present model. One sample (S6) was excluded due to visible damage.

Table 1- Thickness of the samples excised from the central tendon of the right and left sides of the diaphragm.

| Sample number | Thickness value (mm) | Sample number | Thickness value (mm) |
|---|---|---|---|
| Right side | | Left side | |
| T1R | 0.76 | T1L | 0.82 |
| T2R | 0.99 | T2L | 0.76 |
| T3R | 1.00 | T3L | 0.73 |
| T4R | 1.18 | T4L | 0.97 |
| T5R | 0.90 | T5L | 0.94 |
| T6R | Damaged sample | T6L | 1.17 |

Each sample was positioned on a stainless-steel washer and prevented from falling by spreading a fine layer of medical Vaseline between the sample and the washer. The samples were kept hydrated in the saline solution during the preparation procedure. A random pattern was applied to the sample just before the test using very fine carbon particles, Fig. (1-c). Spray paints and alcohol-based markers were avoided because they alter the mechanical response of soft biological tissues. Methylene blue and India ink were also tested but were not retained due to insufficient adhesion on hydrated tendon surfaces.

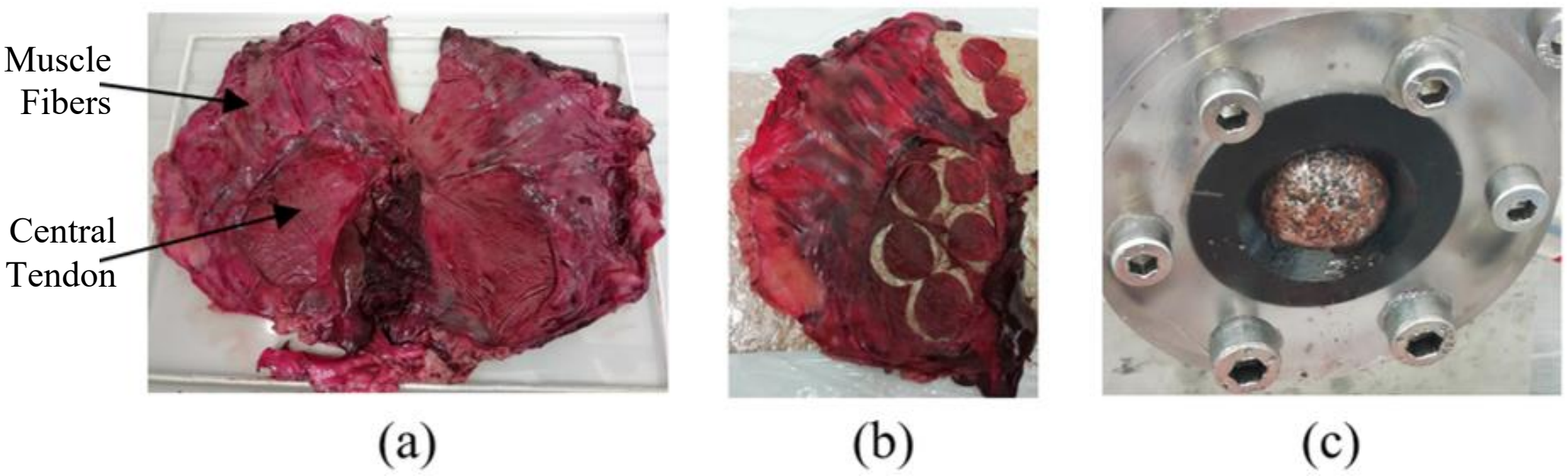

Fig. 1 –(a) The entire diaphragm. (b) Half of the diaphragm with the dissected samples. (c) Sample with the applied speckles mounted on the test setup.

*2.2 Experimental setup and error quantification:*

*2.2.1 Experimental setup*

A custom bulge-test device was developed to enable controlled inflation of the diaphragm samples while allowing simultaneous stereo-DIC acquisition, as shown in Fig. (2). It consisted of three main parts: a rigid base supporting two linear rails for positioning the cameras and the inflation device, a syringe–pusher pressurization system driven by a NEMA 23 stepper motor controlled by a Raspberry Pi 3, and two AlliedVision ® Pike F-421B digita cameras (2048 × 2048 pixels) equipped with Fujinon® HF16HA-1B lenses.

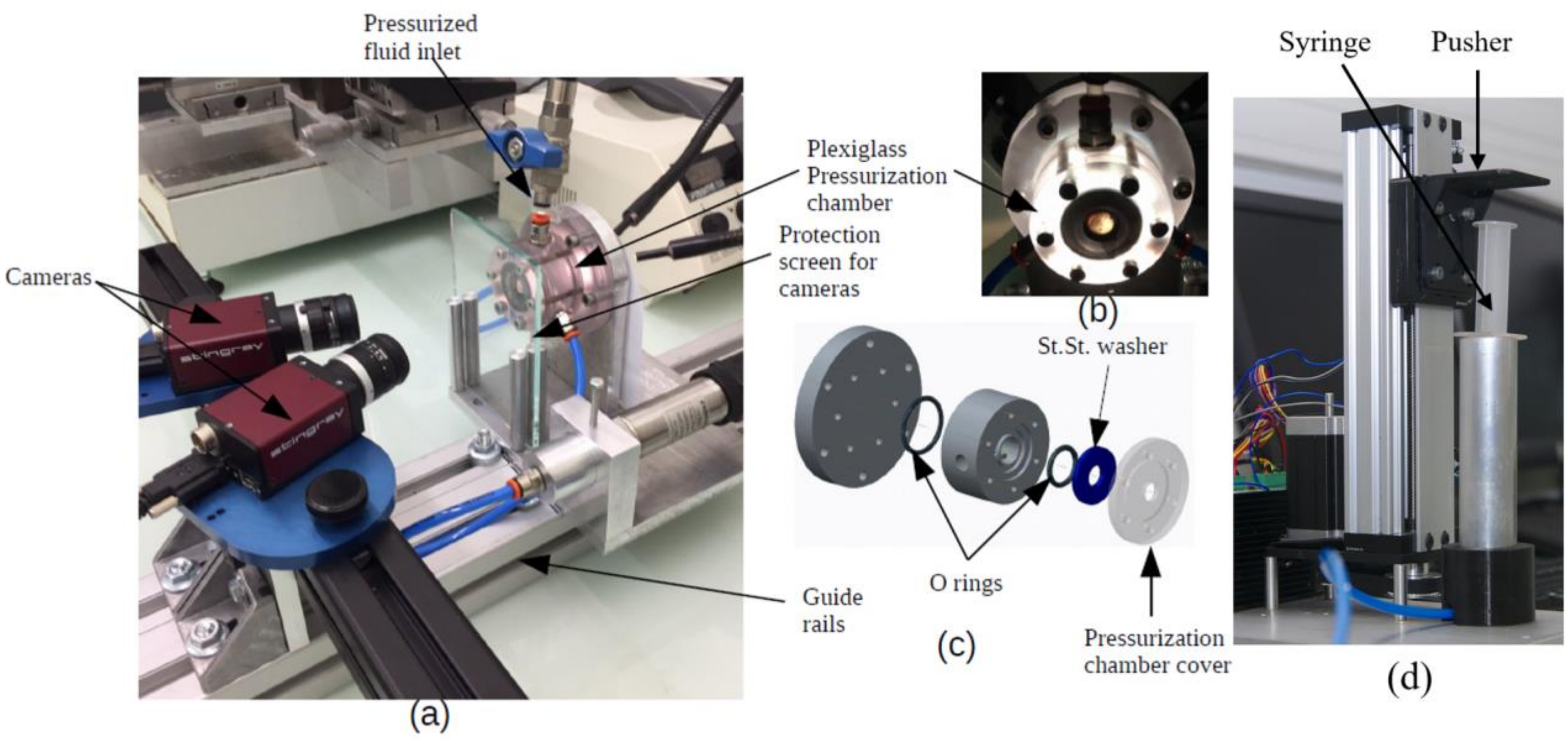

Fig. 2 - Bulge test setup: (a) Setup overview. (b) Plexi-glass pressurization chamber. (c) Schematic form of the chamber dismounted. (d) Syringe pusher device.

The pressurization chamber was machined from transparent plexiglass to allow the use of backlighting, which enhances the visibility of carbon speckles on the translucent tissue and prevents glare resulting from surface hydration. The chamber includes two inlets: one for saline infusion and one for a 0–10 bar Ashcroft® KXD pressure transducer. A protective glass shield was mounted in front of the sample to protect the optical system in case of rupture.

Each specimen was clamped between a stainless-steel washer and an O-ring–sealed cover, Fig. (2). The chamber was filled with saline solution at room temperature. Initial tests performed using compressed air resulted in rapid tissue dehydration; therefore, all subsequent experiments used 0.9% saline (9g of NaCl per 1000 mL of distilled water) as the pressurizing fluid to maintain hydration throughout inflation.

### *2.2.2 Error quantification.*

Because bulge-test–based constitutive identification is highly sensitive to measurement accuracy, several potential sources of uncertainty were examined. Speckle-pattern quality was assessed in the frequency domain, confirming that all patterns exhibited sufficient spatial-frequency richness for reliable correlation. Illumination stability was verified by monitoring grayscale histograms while maintaining constant backlighting and fixed camera exposure settings. Despite specimen-to-specimen biological variability, strain fields showed consistent spatial trends, indicating that correlation noise did not dominate the measurements.

To further evaluate robustness, synthetic bulge-test data were generated from analytical membrane solutions and corrupted with controlled geometric perturbations (up to ±20% on the out-of-plane displacement field). The identified parameters varied by less than 2% in these tests, demonstrating that the identification procedure is highly insensitive to realistic levels of measurement noise.

Detailed procedures and numerical sensitivity analyses are provided in Appendix A - DIC Uncertainty Quantification**.** Together, these results confirm that DIC-related uncertainties do not significantly affect the stretch values used for constitutive modelling.

## *2.3 Experimental procedure:*

Before each test, the mounted specimen was brought to an initial low pressure of approximately 0.05 bar. This preliminary loading step allowed the tissue to unfold and ensured that a clear reference image could be captured for stereo-DIC processing. Because the specimen was firmly clamped between the washer and the cover, no preconditioning cycles were applied. Previous experiments conducted on bovine cornea, porcine sclera, bovine sclera, and human skin tissue [9,18,40,41,42] have reported negligible influence of preconditioning on the inflation response under similar boundary conditions, justifying this choice.

Inflation was performed by gradually injecting the physiological serum into the pressurization chamber using the motor-driven syringe pump. As the pressurized fluid entered, it forced the diaphragm tissue to deform into a spherical dome shape. Pressure was increased in increments of approximately 0.03 bar. After each increment, a relaxation period of about 60 seconds was imposed to allow the sample to reach a quasi-steady state before image acquisition. This protocol reduced transient effects that could degrade DIC measurements and is consistent with procedures reported in the literature for soft tissue bulge testing [7]. The maximum pressure reached during testing was approximately 0.7 bar. Pressures beyond this value were avoided because higher loading resulted in fluid percolation through the tissue, which disrupted the carbon speckle pattern and prevented reliable DIC tracking. All tests were conducted at room temperature. For each pressure level, synchronized images from the two cameras and the corresponding pressure signal were recorded using Vic-Snap 2009 (Correlated Solutions Inc., Columbia, SC). These data were later processed in VIC-3D® to obtain the full-field displacement and strain distributions needed for subsequent stretch computation and constitutive parameter identification.

## *2.4 DIC measurements and calibration*

In the current study, digital image stereo-correlation (stereo-DIC) was adopted to assess the strain distribution. This method enables the mapping of three-dimensional deformation with high spatial and temporal resolution. For effective implementation, stereo-DIC necessitates a surface with contrasting dark and light features that deform along with the material. To achieve this requirement, a speckle contrast pattern was generated on the diaphragm samples.

Before conducting the inflation tests, an essential calibration phase is undertaken to convert pixel position information into metric data. A camera's intrinsic parameters, such as its main point, focal length, and distortion coefficients, are established within a defined reference frame. To determine these parameters, a test pattern featuring a 12x9 grid of points spaced 2mm apart is utilized. This pattern is positioned in various orientations within the camera's viewing volume. Calibration entails moving the pattern along the x and y axes, as well as rotating it around the three axes. System calibration was done according to the specifications outlined by the VIC-3D system, yielding a reconstruction error below 0.03 pixels, in agreement with high-quality DIC systems reported in the literature. Because the stretch determination used for constitutive identification is sensitive to measurement quality, a focused uncertainty analysis was performed. Only the main conclusions are reported here, while full details (speckle-pattern Fourier evaluation, illumination stability, and numerical noise-sensitivity tests) are provided in Appendix A**.**

Each image collected in this way makes it possible to match the actual positions of the checkerboard intersections in space with their pixel coordinates in the CCD plane. This correlation allows the determination of the camera's intrinsic parameters. The so-called "extrinsic" parameters define the translation and rotation to move from the retinal marker of camera 1 to that of camera 2 and are determined from an image (or more) of the same test pattern. This calibration data empowers the calculation of 3D point coordinates in space when the pixel locations of camera 1 and camera 2 viewing that point are known. The field of view for our setup was a 14mm diameter disk, representing the exposed sample diameter observed by the cameras. The out-of-plane inflation of the capsule reached approximately 5mm [43].

Throughout the inflation process, images were acquired at intervals of 3 seconds. The VIC 3D ® stereo-correlation software [19,44] has been used to determine the displacement field on the surface of the diaphragmatic tendon samples. This was achieved by reconstructing the 3D spherical shape of the inflated sample captured in each image and reporting the current position of each speckle applied to the surface. Indeed, the deformation of the random speckle pattern reflects the surface deformation of the samples. The data obtained from VIC-3D, regarding the initial and instantaneous positions of each speckle within the area of interest during the test, was integrated into a computational code using Python. Circumferential and meridional curvature radii were computed, enabling the calculation of tissue stretch ratios in both directions, as elaborated in the following section.

It must be emphasized that the initial shape of the diaphragm samples exhibited inherent deformation. This deformation occurred during the process of securing and tightening the tissue within the setup grips. The sample being incompressible, its compression at the edges involved the displacement of the border material toward the center, forcing it to buckle. Acknowledging this, it was imperative to account for this initial deformation to accurately identify the parameters. The $F$, strain gradient tensor, calculated from the stretches, was considered to be $F_{total}$, while the deformed initial state was represented as $F_0$. Hence, there exists a term $F_{id}$ (F identification) to be derived from this relationship:

$$F_{total} = F_0 F_{Id} \Leftrightarrow F_{Id} = F_0^{-1} F_{total} \quad (2.1)$$

Provided that the tensors $F$ are represented by diagonal matrices, this leads to:

$$\lambda_{i_{Id}} = \frac{\lambda_i}{\lambda_{i_0}}, i = 1, 2, 3 \quad (2.2)$$

λ is the stretch which corresponds to the ratio of the current (deformed) length "l" to the reference (undeformed) length "L".

### *2.5 Approximation of the deformed specimen to a spherical cap*

To compute the curvature and stretch fields from the reconstructed 3D geometry, the inflated diaphragm was approximated as a spherical cap. This assumption was adopted for geometric convenience only and does not imply that the tissue behaves as an isotropic material. As illustrated by the microstructural observations, Fig. (3) and confirmed by the constitutive parameter identification, the central tendon exhibits pronounced anisotropic features. The spherical approximation was therefore used exclusively to derive geometric quantities, curvatures and stretches, while the material behavior itself was modeled later using a transversely isotropic hyperelastic formulation. Under this approximation, the meridional curvature is obtained directly from the DIC-measured profile, whereas the circumferential curvature is inferred from axisymmetry. This methodology follows the work of Machado et al. [26], who demonstrated that spherical fitting provides accurate curvature and stretch estimates at the pole when a sufficient number of reliable measurement points is available.

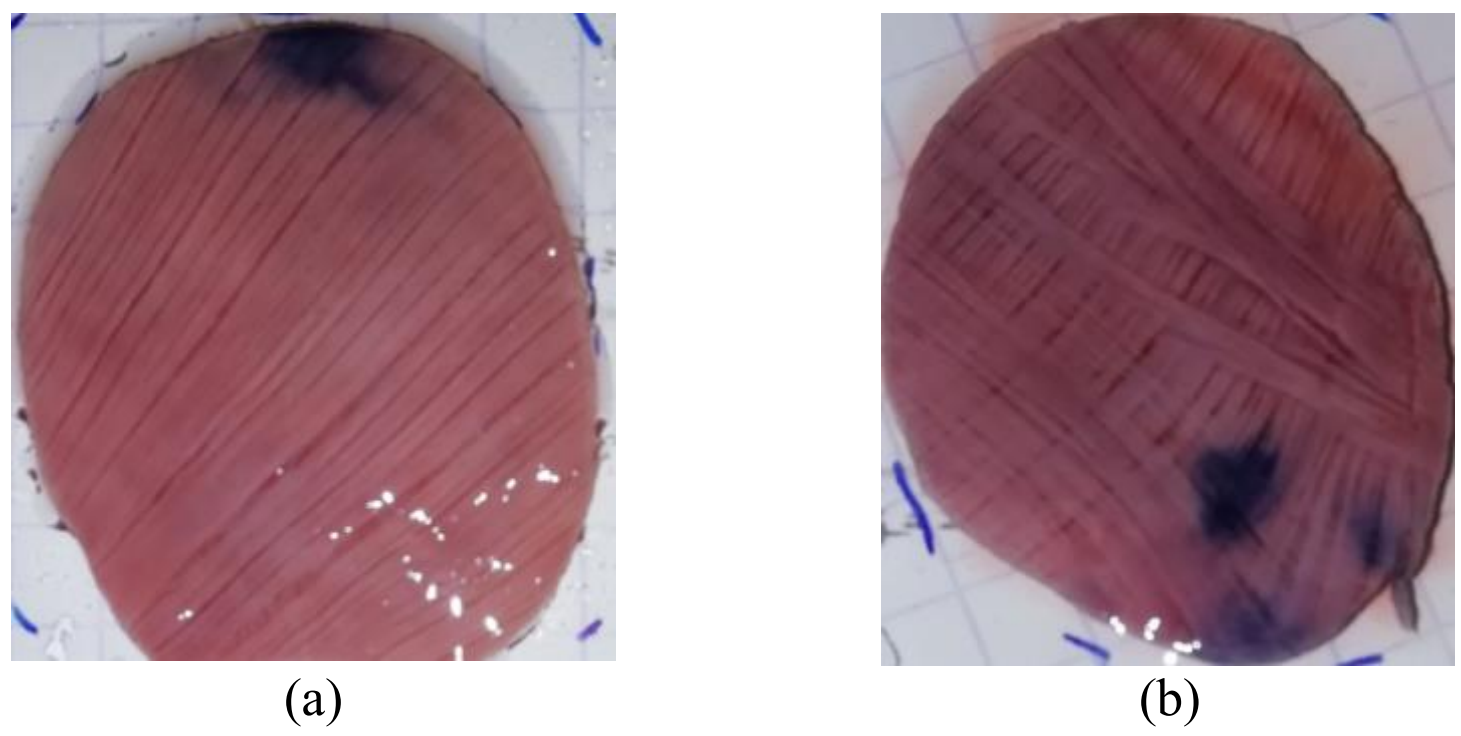

(a) (b)

Fig. 3 - Close view of 2 samples dissected from the diaphragm CT. (a) sample where the fibers have privilege orientation. (b) sample where the fibers are oriented in 2 directions.

Throughout each step of the stereo-correlation process, the inflated capsule is approximated as a sphere with a radius *r* and a center *C*. For a visual representation, please refer to Fig. (4) which details the description of the inflated capsule.

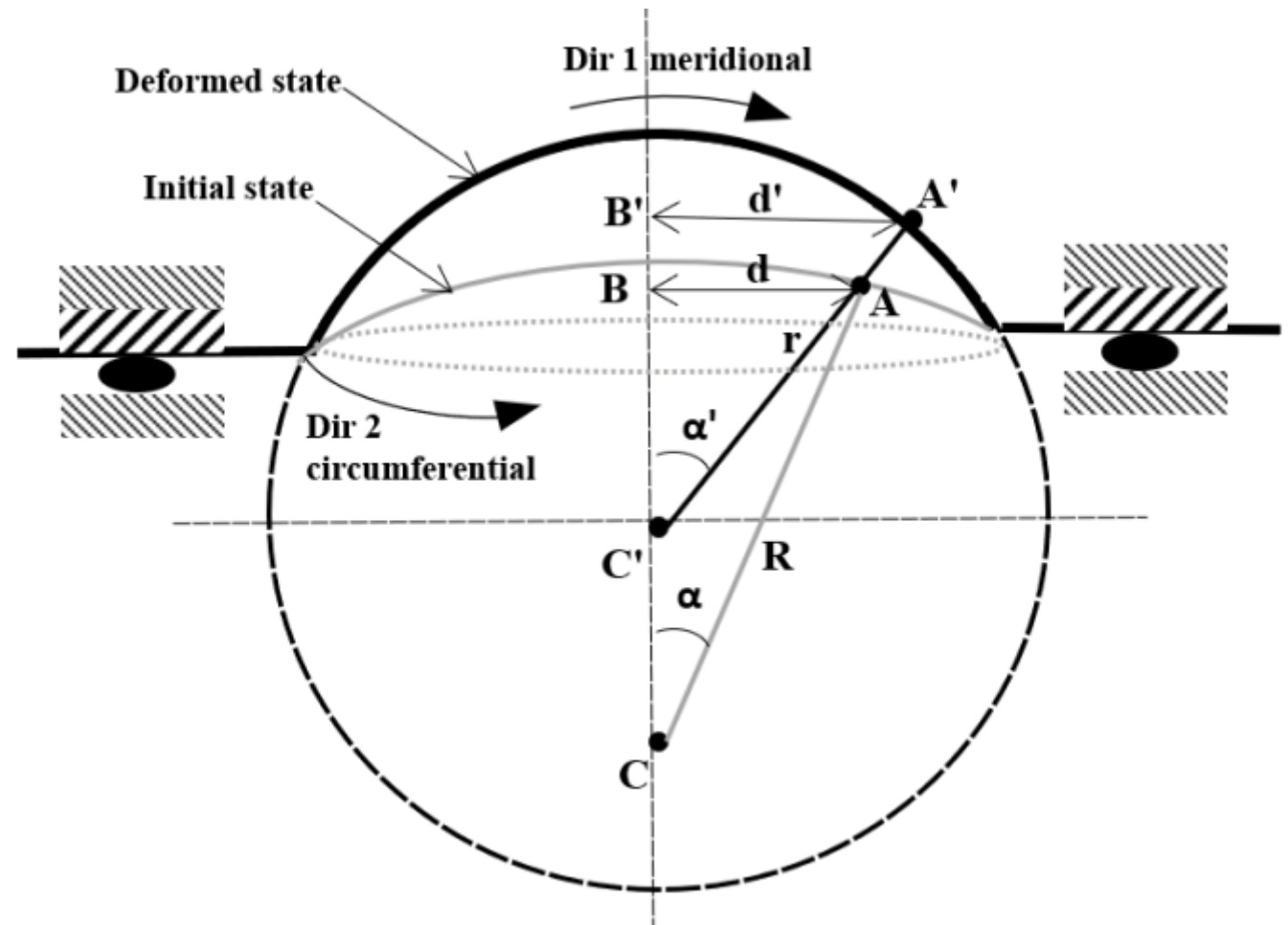

Fig. 4 - The approximated shape of the inflated sample.

To determine the sphere radius $r$ and center $C'$, we started from the general form of an ellipsoid equation, which comes from the equation of a quadric surface:

$$ax^2 + by^2 + cz^2 + 2dxy + 2exz + 2fyz + 2gx + 2hy + 2iz + j = 0 \tag{2.3}$$

Including the conditions of a sphere centered in C= (0,0,$z_c$), we have the following simplification:
$a = b = c = 1$
$d = e = f = g = h = 0$
This gives us the following simpler equation:

$$x^2 + y^2 + z^2 + 2iz + j = 0 \tag{2.4}$$

The ellipsoid equation may be also written in matrix form:

$$(x, y, z)^T \boldsymbol{Q}(x, y, z) = 1 \tag{2.5}$$

The eigenvectors and eigenvalues of the positive definite 4x4 matrix Q define the axis and radii, respectively, of the ellipsoid. In the case of a sphere, the orientation of the axis does not matter, However, the eigenvalues hold significance as their inverses denote the radii, which in our case are equal.

According to Brunon et al, [19], $r$ and $R$ allow the determination of the meridional and circumferential stretch ratios ($\lambda_1$ and $\lambda_2$ respectively), where $R$ represents the radius of the flat specimen. Notably, the initial state of the capsule was not flat, and the thickness of the specimen was assumed to be homogeneous. Referring to the description of the inflated diaphragm in Fig. (3), we can determine the mean stretch ratio as follows:

$$\begin{cases} \lambda_1 = \dfrac{1}{n}\displaystyle\sum_{i=1}^{n} \dfrac{\alpha_i' R_i'}{\alpha_i R_i} \\ \lambda_2 = \dfrac{1}{n}\displaystyle\sum_{i=1}^{n} \dfrac{d'}{d} \end{cases} \tag{2.6}$$

where $\alpha_i$ and $R_i$ denote the angular and radial coordinates in the reference state, and $\alpha_i'$ and $R_i'$ their deformed counterparts. The quantities $d$ and $d'$ represent the circumferential arc lengths before and after deformation, Fig. (4).

## 3. Modelling of the hyperelastic behavior

### *3.1 Incompressible isotropic hyperelastic models*

To describe the mechanical response of the diaphragmatic central tendon under equibiaxial inflation, a hyperelastic constitutive framework based on strain-energy density functions was adopted. In this first step, incompressible isotropic hyperelastic models were considered as phenomenological baselines, motivated by the macroscopically axisymmetric deformation observed at the pole during bulge tests. These models are not intended to represent the underlying fibrous microstructure of the tissue, but rather to provide reference descriptions for comparison with anisotropic formulations introduced subsequently.

Under the assumptions of isotropy and incompressibility, the Cauchy stress tensor was formulated using the Rivlin–Ericksen expression, which links stress to the strain invariants of the deformation tensor [7]:

$$\boldsymbol{\sigma} = -p\boldsymbol{Id} + 2\frac{\partial\psi}{\partial I_1}\boldsymbol{B} - 2\frac{\partial\psi}{\partial I_2}\boldsymbol{B}^{-1} \tag{3.1}$$

where $p$ is the arbitrary hydrostatic pressure, **Id** is the identity tensor, and **B** is the left Cauchy-Green strain tensor: $\mathbf{B} = \mathbf{F}\mathbf{F}^{\mathbf{T}}$, where **F** is the strain gradient tensor. The principal invariants $I_1$, $I_2$ and $I_3$ are defined by:

$$I_1 = tr(\boldsymbol{B}), \tag{3.2}$$

$$I_2 = \frac{1}{2}[I_1^2 - tr(\boldsymbol{B}^2)], \tag{3.3}$$

$$I_3 = det(\boldsymbol{B}) = 1\ (incompressibility). \tag{3.4}$$

The material response coefficients can equivalently be expressed in terms of the principal stretches $\lambda_1$, $\lambda_2$, $\lambda_3$:

$$2\frac{\partial\psi}{\partial I_1} = \frac{1}{\lambda_1^2 - \lambda_2^2}\left(\frac{\lambda_1^2 + \lambda_3^2}{\lambda_1}\frac{\partial\psi}{\partial\lambda_1} - \frac{\lambda_2^2 + \lambda_3^2}{\lambda_2}\frac{\partial\psi}{\partial\lambda_2}\right) \tag{3.5}$$

$$-2\frac{\partial\psi}{\partial I_2} = \frac{1}{\lambda_1^2 - \lambda_2^2}\left(\frac{1}{\lambda_1}\frac{\partial\psi}{\partial\lambda_1} - \frac{1}{\lambda_2}\frac{\partial\psi}{\partial\lambda_2}\right) \tag{3.6}$$

We successively consider four incompressible isotropic hyperelastic models, namely, Neo-Hookean, Mooney-Rivlin, Yeoh, and Fung models. The first three models come from rubber-like materials studies and are characterized by a polynomial strain function. The strain function is as follows:

$$\psi_{polynomial} = \sum_{i=0}^{ni}\sum_{j=0}^{nj} C_{ij}(I_1 - 3)^i(I_2 - 3)^j \tag{3.7}$$

Where $C_{ij}$ are the materials parameters. The Neo-Hookean model is obtained when $n_i = 1$ and $n_j = 0$, the two-parameter Mooney-Rivlin model is characterized by $n_i = 1$, $n_j = 1$, and $C_{11} = 0$ and Yeoh model corresponds to $n_i = 1$ and $n_j = 1$. This leads to:

$$\psi_{Neo-Hookean} = C_{10}(I_1 - 3) \tag{3.8}$$

$$\psi_{Mooney-Rivlin} = C_{10}(I_1 - 3) + C_{01}(I_2 - 3) \tag{3.9}$$

$$\psi_{Yeoh} = C_{10}(I_1 - 3) + C_{20}(I_1 - 3)^2 \tag{3.10}$$

The Fung model was specifically developed to capture the response of soft tissues with a high content of collagen fibers [45, 46] and given by an exponential strain function:

$$\psi_{Fung} = \frac{a}{2b}\left[e^{b(I_1-3)} - 1\right] \tag{3.11}$$

where a and b are the materials parameters.

Incorporating the strain energy density functions into Eq. (3.1) allows the derivation of the expressions of the Cauchy stress tensors for the four distinct models:

$$\boldsymbol{\sigma}_{Neo-Hookean} = -p\boldsymbol{Id} + 2C_{10}\boldsymbol{B} \quad with\ p = 2C_{10}\lambda_3 \tag{3.12}$$

$$\boldsymbol{\sigma}_{Mooney-Rivlin} = -p\boldsymbol{Id} + 2C_{10}\boldsymbol{B} - 2C_{01}\boldsymbol{B}^{-1} \quad with\ p = 2C_{10}\lambda_3 - 2C_{01}\lambda_3 \tag{3.13}$$

$$\boldsymbol{\sigma}_{\boldsymbol{Yeoh}} = -p\boldsymbol{Id} + 2\big(C_{10} + 2C_{20}(I_1 - 3)\big)\boldsymbol{B} \quad with\ p = 2\big(C_{10} + 2C_{20}(I_1 - 3)\big)\lambda_3 \tag{3.14}$$

$$\boldsymbol{\sigma}_{\boldsymbol{Fung}} = -p\boldsymbol{Id} + \frac{a}{2}e^{b(I_1-3)}\boldsymbol{B} \quad with\ p = \frac{a}{2}e^{b(I_1-3)}\lambda_3 \tag{3.15}$$

Note that, because of the material incompressibility, det(B) = 1, the stretch along the thickness, $\lambda_3$, is a function of the two other stretches, $\lambda_1$ and $\lambda_2$. Furthermore, since the diaphragm is assimilated to a thin membrane, a state of plane stress is considered, which implies that $\sigma_{33}$ is negligible.

*3.2 Transversely isotropic hyperelastic model (Humphrey–Yin)*

To account for the effective anisotropic response associated with the layered structure of the diaphragmatic tendon, a transversely isotropic hyperelastic formulation based on the Humphrey-Yin framework was adopted. Although this formulation is often interpreted in terms of explicit fiber reinforcement, it can equivalently be viewed as an invariant-based transversely isotropic hyperelastic model. In the present work, this interpretation is used to describe a membrane that is isotropic within its plane while exhibiting a distinct mechanical response through its thickness [47].

The tissue was modeled using a transversely isotropic Humphrey–Yin formulation, with the axis of symmetry aligned with the membrane thickness direction. In this framework, the anisotropic invariant is defined as $I_4 = \lambda_3^2$, where the preferred material direction coincides with the membrane's normal. The material parameters $\mu$, $k_1$, and $k_2$ govern the isotropic matrix stiffness and the magnitude and nonlinearity of the transversely isotropic contribution, respectively. By construction, this formulation preserves isotropy within the membrane plane and does not introduce an explicit in-plane fiber family. Instead, the anisotropic contribution captures effective through-thickness stiffening associated with the stratified structure of the tendon, allowing distinct mechanical behavior in the thickness direction while maintaining an isotropic in-plane response. The improved agreement with experimental observations should therefore be interpreted as a manifestation of transverse stiffening governing the inflation response, rather than evidence of oriented fiber reinforcement within the membrane plane. The strain–energy density function was expressed as the sum of an isotropic matrix contribution and a transversely isotropic contribution,

$$\psi = \psi_{iso}(I_1) + \psi_{ti}(I_4), \qquad (3.16)$$

Where $I_1$ is the first invariant of the right Cauchy–Green deformation tensor $\mathbf{C}$, and $I_4$ is the invariant associated with transverse isotropy. The isotropic contribution was defined as:

$$\psi_{\text{iso}} = \frac{\mu}{2}(I_1 - 3), \qquad (3.17)$$

where $\mu$ denotes the shear modulus of the isotropic matrix.

The transverse invariant was defined as: $I_4 = \mathbf{a}_0 \cdot \mathbf{C}\mathbf{a}_0$, (3.18)

where $\mathbf{a}_0$ denotes the preferred material direction in the reference configuration. In the present formulation, this direction was aligned with the membrane normal. Under the incompressibility assumption, this yields $I_4 = \lambda_3^2$, where $\lambda_3$ is the stretch in the thickness direction.

The transversely isotropic contribution was expressed as:

$$\psi_{\text{ti}} = \frac{k_1}{2k_2}\left[\exp(k_2(I_4 - 1)^2) - 1\right] \qquad (3.19)$$

where $k_1$ and $k_2$ are material parameters controlling the magnitude and nonlinearity of the transverse stiffening response.

With this formulation, isotropy is preserved within the membrane plane by construction, and the anisotropic contribution accounts solely for a through-thickness effect associated with the stratified structure of the tendon.

For incompressibility: $$\lambda_3 = \frac{1}{\lambda_1\lambda_2} \quad (3.20)$$

The Cauchy stress in the direction 1 becomes:

$$\sigma_1 = 2\lambda_1[\mu\lambda_1 + 2k_1\lambda_1(I_4 - 1)exp(k_2(I_4 - 1)^2)]. \quad (3.21)$$

*3.3 Equilibrium equation (EE) for a membrane*

Among the fundamental relations used in our study is the equation of equilibrium for a thin membrane in the direction tangential to the surface. This relation connects experimental pressure to the meridional and circumferential stresses. According to Hill 1950 [8], this equation may be written as follows:

$$\frac{\sigma_1}{\rho_1} + \frac{\sigma_2}{\rho_2} = \frac{p}{h} \quad (3.22)$$

$\rho_1$ and $\rho_2$ are the meridional and circumferential radius of curvature; $\sigma_1$ and $\sigma_2$ are the meridional and circumferential stresses, $p$ is the experimental pressure, and $h$ is the deformed thickness. Generally, the radii are not equal except at the pole of the inflated specimen unless the bulge happens to be a spherical cap. For isotropic material, we considered the deformed specimen to have a spherical shape, $\rho_1 = \rho_2 = r$. Thus, the relation (3.16) becomes:

$$\Rightarrow \frac{1}{r}(\sigma_1 + \sigma_2) = \frac{p\lambda_1\lambda_2}{H} \quad (3.23)$$

However, we didn't measure the deformed thickness, but assuming incompressibility, the deformed $h$ thickness can be calculated from the undeformed thickness $H$ and stretch ratios such as:

$$h = H\lambda_3 = \frac{H}{\lambda_1\lambda_2} \quad (3.24)$$

If we rewrite the equilibrium equation (3.17) and replace the $\sigma_i$ with the stress from the Yeoh model, we will be able to put the equilibrium equation in the form of $Ax = B$ and use the least square method for the optimization. On these terms:

$$A = \frac{2}{r}\left(\lambda_1^2 + \lambda_2^2 + \frac{2}{\lambda_1^2\lambda_2^2} \qquad 2(I_1 - 3)\left(\lambda_1^2 + \lambda_2^2 + \frac{2}{\lambda_1^2\lambda_2^2}\right)\right)$$

$$B = \frac{p\lambda_1\lambda_2}{H}$$

$$x = (C_{10} \quad C_{20})^T$$

The same approach was adopted for the other three hyperelastic models. Hence the $C_{ij}$ of different models was calculated. Using Python, the following relation was resolved:

$$\frac{\sigma_1 + \sigma_2}{r}\left(\frac{H}{\lambda_1\lambda_2}\right)x = p \quad (3.25)$$

Where $p$ is the experimental pressure in MPa and the term $\frac{\sigma_1+\sigma_2}{r}\left(\frac{H}{\lambda_1\lambda_2}\right)$ represents the pressure calculated from the models.

*3.4 Identification of the material parameters*

The other step in the material parameters identification process involves minimizing the gap between the experimental and model-calculated stresses. To achieve this, we used different algorithms. Firstly, for the polynomial models, we used a constrained Sequential Least SQuares Programming (SLSQP). Secondly, for the exponential models (Fung and Humphrey-Yin), we used a bounded Limited-memory Broyden–Fletcher–Goldfarb–Shanno (L-BFGS-B) algorithm. The inclusion of constraints was crucial to ensure the numerical robustness of the strain energy [48]. We adopted the following constraints:

- Neo-Hookean: $C_{10} \geq 0$
- Yeoh: $\min(2C_{10} + 4C_{20}(I-3) \geq 0$
- Mooney-Rivlin: $C_{10} + C_{01} \geq 0$
- Exponential (Humphrey-Yin): $a \ and \ b \geq 0 \ and \ k_1 \ and \ k_2 \ \geq 0$

We chose to minimize the function ($C_{ij}$, a, b), which corresponds to the sum of the squared gap within the equilibrium for each level of pressure $P_i$:

$$\mathrm{f}\left(C_{ij}, \mathrm{a}, \mathrm{b}\right) = \sum\left((\sigma_1 + \sigma_2) - \left(\frac{p_i \lambda_1 \lambda_2}{H}\right)\right)^2 \quad (3.26)$$

Where $C_{ij}$, a and b are the material parameters found in the considered energy density. $\sigma_1$ and $\sigma_2$ are the stresses for the current set of parameters and the current pressure. The optimization was done using Python and more precisely the *minimize* function of the *optimize* module of Scipy. We kept the expressions of $\sigma_1$ and $\sigma_2$ to verify the assumption of an isotropic diaphragmatic membrane.

## 4. Numerical validation

To validate the approach of approximation of the inflated sample shape to a spherical cap, we proposed generating virtual bulge test data using the Neo-Hookean model, as illustrated in Fig. (5-a). We aim to evaluate the error between the imposed material parameter $C_{10} = 0.5$ and the calculated parameter using the proposed method. This involves creating data points that form a discrete hemisphere, to which we assign specific pressures. Consequently, we can derive stress values from these points and substitute them into the equilibrium equation to calculate the parameter value. The error, in this context, represents the disparity between the input and output values:

$$Error = \left|\frac{C_{10theo} - C_{10cal}}{C_{10cal}}\right| \quad (4.1)$$

It is not likely to get perfect shape from the stereo-correlation technique, so it was needed to check the robustness of the method against the noisy data. We followed the same protocol, but we added some noise to the generated data, Fig. (5-b). To do so we simply multiplied the z values by a random number between 0.8 and 1.2 obtained using a random uniform function (Python function numpy.randon.rand). The resulting error in the identified material parameters remained below 2%, demonstrating the robustness of our approach.

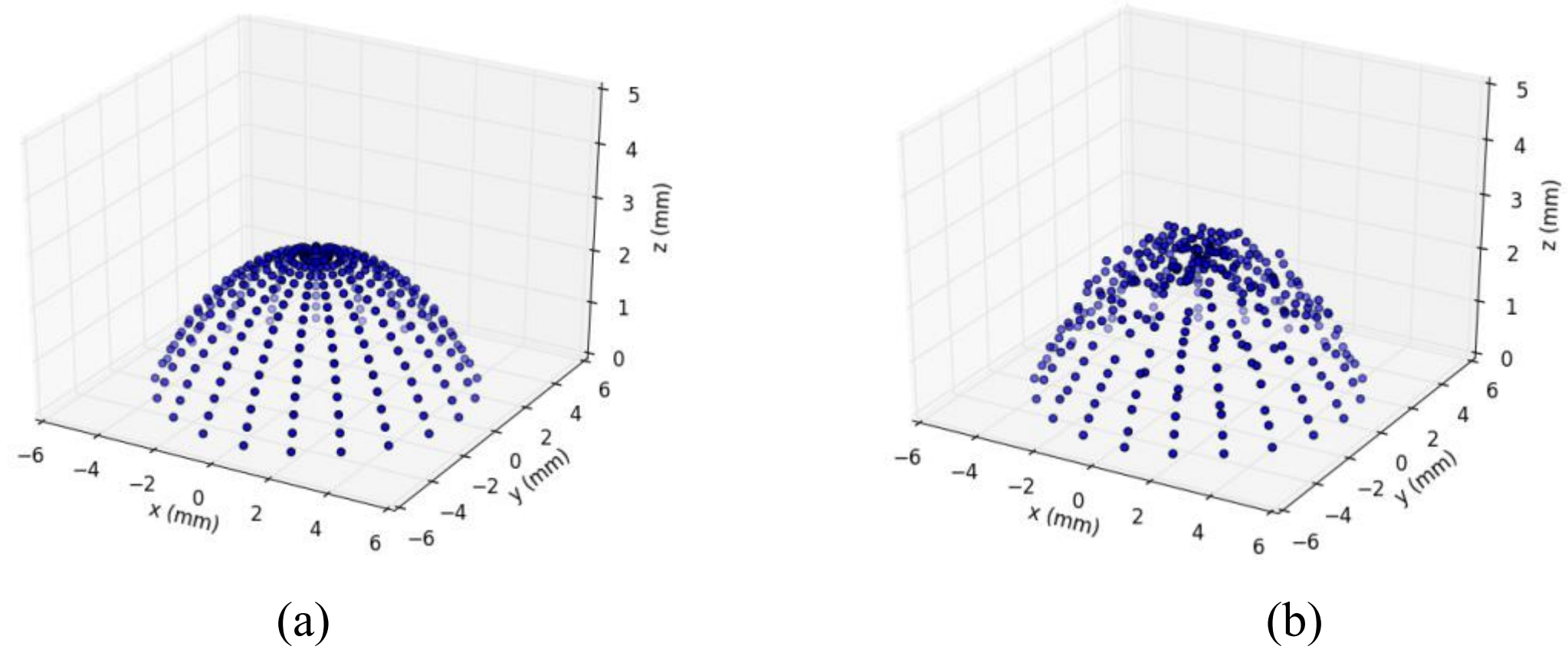

Fig. 5 - Virtual experimental data: (a) perfect spherical cap. (b) data with noise.

## 5. Results and discussion

A sample of the results obtained using stereo-DIC is shown in Fig. (6). Fig. (6-a) shows the stretch distribution on the inflated tissue surface. Fig. (6-b) shows the evolution of the shape of the inflated capsule plotted in the X-Z plane at different pressure values. The disruption of data, shown in Fig. (6-b), is due to the reflection of light on the wet tissue.

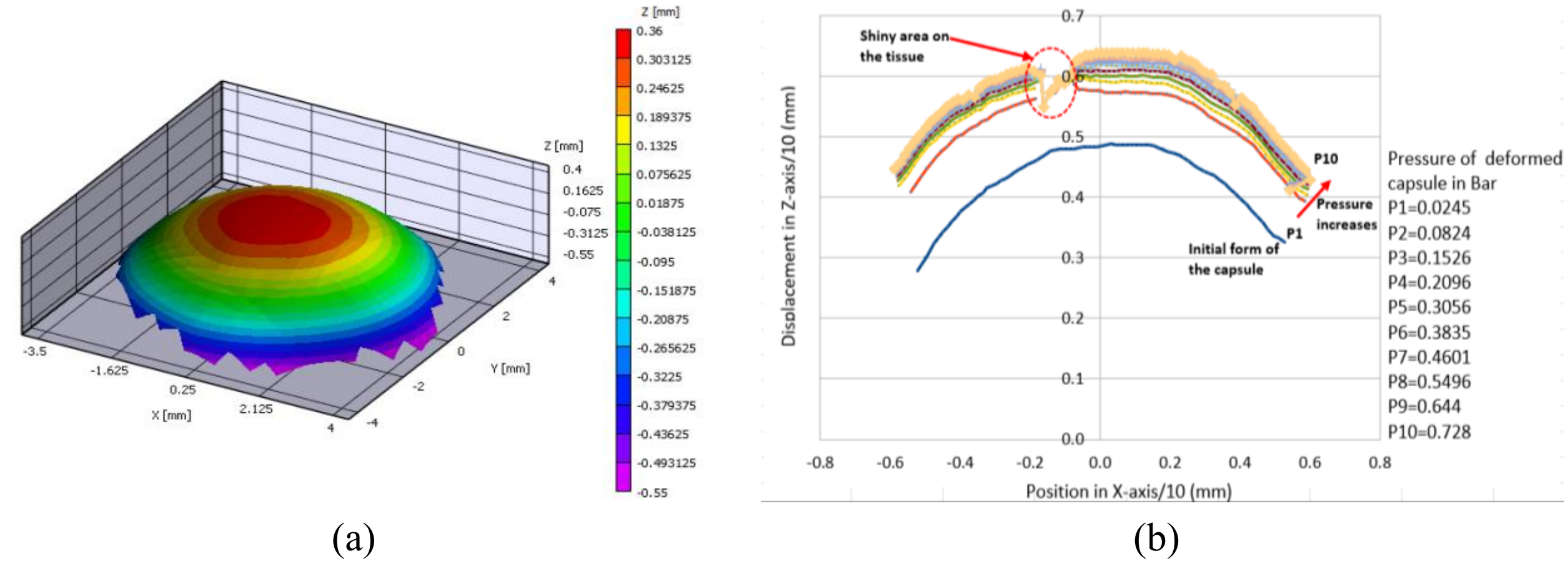

Fig. 6 – (a) Stretch distribution on the surface of the inflated tissue after stereo-correlation. (b) Example of the evolution of the external shape of the capsule for one sample extracted from the right half of the diaphragm. Displacement in X-Z plane for different pressure values in bars.

The relations between the pressure and the meridional and circumferential stretches ratios obtained from bulge tests for the eleven samples extracted from both halves of the diaphragm are shown in Fig. (7). The visible noise on the curves results from the incremental pressure application, consisting of steps of 0.03 bar, separated by relaxation intervals. These curves exhibit the expected classically hyperelastic behavior typical of soft biological tissues. A noteworthy observation is the considerable variability among results obtained from different samples despite their common origin within the same animal. This discrepancy cannot be attributed to the sample's origin or handling protocol. Instead, it likely stems from the non-uniform structure of the tendon, evident in Fig. (4), showcasing diverse fibers orientations.

This significant dissimilarity in sample behavior points to the anisotropic nature of the tendon. Comparing $\lambda_1$ and $\lambda_2$ in Fig. (7), it's evident that the circumferential direction exhibited less stretch than the meridional direction at the same pressure level. Nonetheless, two distinct behaviors emerge: samples 3, 4, and 5 from the left half and sample 4 from the right half of the diaphragm exhibit nearly identical stretch values in both directions, possibly indicating fibers oriented in two different directions, as depicted in Fig. (4-b). Conversely, the remaining samples show a considerable difference between $\lambda_1$ and $\lambda_2$ suggesting a predominant fiber orientation, as shown in Fig. (4-a). Table 1 further illustrates variations in sample thickness, potentially indicating multiple layers of fibers bundles oriented in different directions. Additionally, the amount of relaxation appears to correlate with the stretch level, with more noticeable effects at lower stretch levels. Unfortunately, a comparison in terms of failure stresses couldn't be established as the tests ceased when correlation points could no longer be tracked.

To address the issue of experimental uncertainty, we conducted a numerical validation of our identification method. Specifically, we generated synthetic bulge test data with added random noise (±20%) to simulate measurement variability. The resulting error in the identified material parameters remained below 2%, demonstrating the robustness of our approach as illustrated in section 4. The uncertainties associated with 3D Digital Image Correlation (3D-DIC) were summarized in section 1. Finally, the bulge test method itself has been validated in several studies for soft membrane characterization. Jourdan et al. (2023) [23] demonstrated that combining 3D-DIC with analytical modeling provides reliable stress and strain estimations, even under complex deformation conditions.

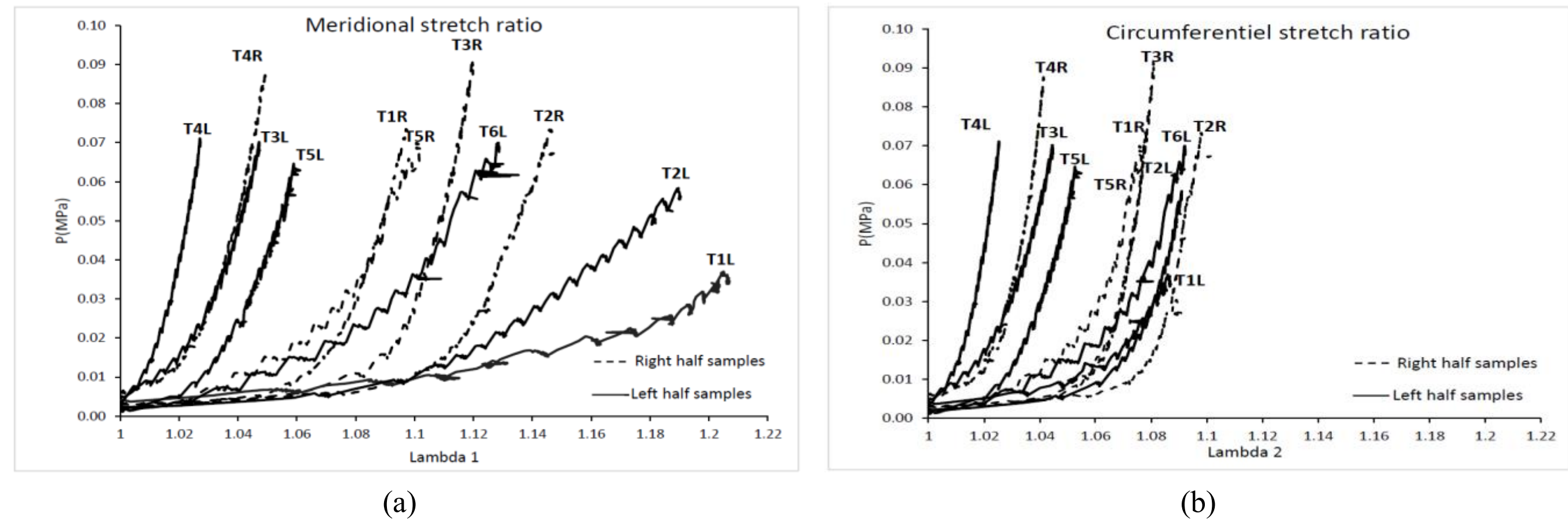

(a) (b)

Fig. 7 - P – stretch ratio curves for experimental data of the samples extracted from the right and left halves of the diaphragm. (a) meridional stretch ratio. (b) circumferential stretch ratio, T: test, R: right, L: left.

*5.1 Incompressible isotropic hyperelastic models*

The relations between pressure and the meridional and circumferential stretch ratios obtained from bulge tests and Yeoh and Fung hyperelastic models for the two halves of the diaphragm are shown in Fig. (8 – 11). The Y-axis represents the pressure values recorded during the tests for the experimental curves. However, it corresponds to the left-hand side term, which is equivalent to $A$ in the form $Ax=B$ of the equation (3.19), for the hyperelastic models. The maximum applied pressure reached approximately 0.7 bar, resulting in peak meridional and circumferential stretch ratios between 1.1 and 1.2. Further increases in stretch ratios were unattainable due to the passage of physiological serum through the tissue samples, which disrupted the carbon particle pattern when higher pressures were applied. As a result, the cameras failed to accurately track surface deformation.

Cesare et al [6] highlight that a hyperelastic Neo-Hookean constitutive model effectively characterizes the mechanical behavior of the central tendon. This is attributed to its high stiffness, which primarily drives the deformation of the diaphragm through muscle contraction. However, our experimental bulge test data revealed an exponential relationship between pressure and stretch in the diaphragm tissue, rendering the Neo-Hookean model inadequate for reproducing the membrane's mechanical behavior. Furthermore, the Mooney-Rivlin model poorly fits the experimental results of the tested samples, indicating the inadequacy of these hyperelastic models, for single-phase incompressible isotropic materials, for describing the behavior of this fibrous biological tissue. Consequently, results obtained from these models were disregarded due to their failure to predict observed behavior. While Yeoh's second-order model displayed improved alignment with the tested samples' behavior, it tended to overestimate the material's non-linear behavior, as indicated in Fig. (8). Ultimately, the Fung model, constructed upon exponential stored energy functions designed for arterial deformation [49, 50], effectively portrayed the mechanical behavior of the hyperelastic material, as shown in Fig. (9) and (11).

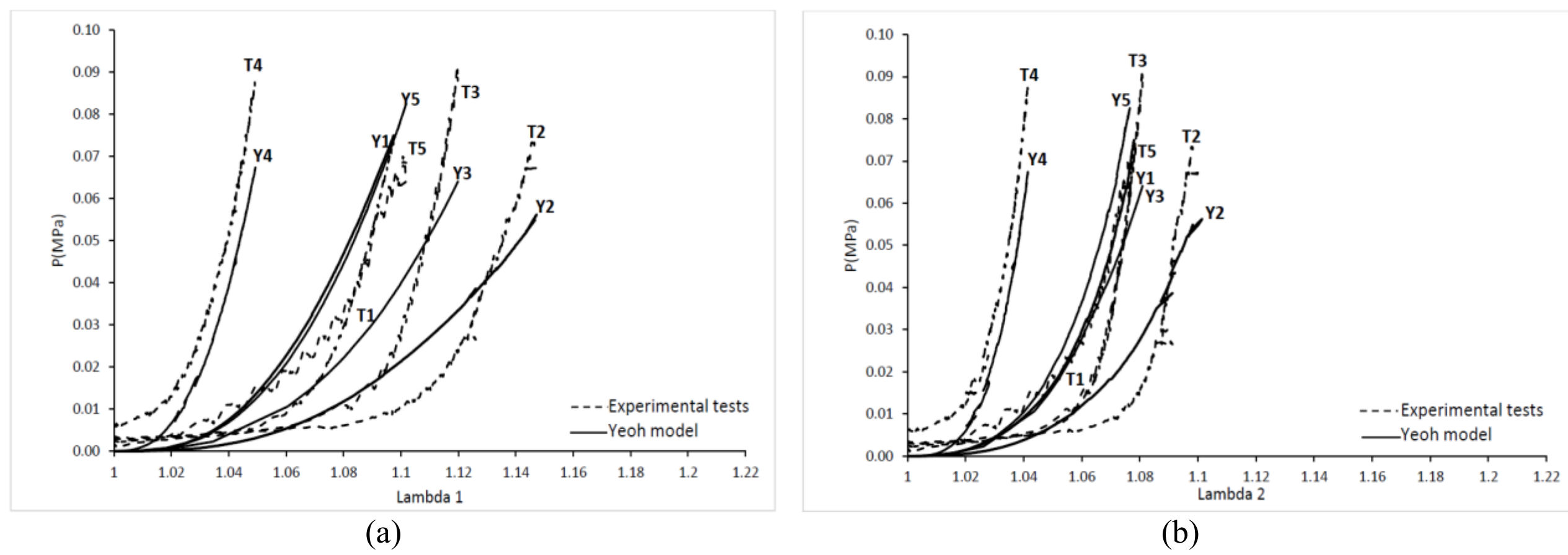

(a) (b)

Fig. 8 - P – stretch ratio curves for experimental data and Yeoh model of the five samples extracted from the right half of the diaphragm. (a) meridional stretch ratio. (b) circumferential stretch ratio. T: test, Y: Yeoh model.

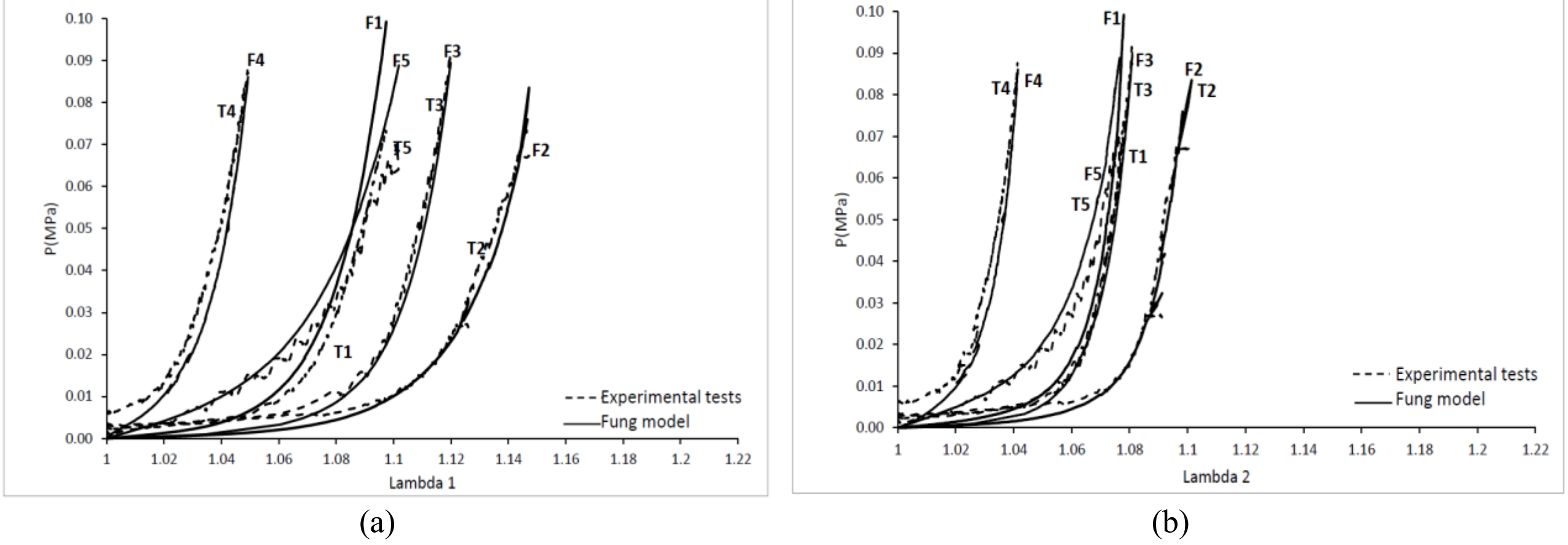

(a) (b)

Fig. 9 - P – stretch ratio curves for experimental data and Fung model of the five samples extracted from the right half of the diaphragm. (a) meridional stretch ratio. (b) circumferential stretch ratio. T: test, F: Fung model.

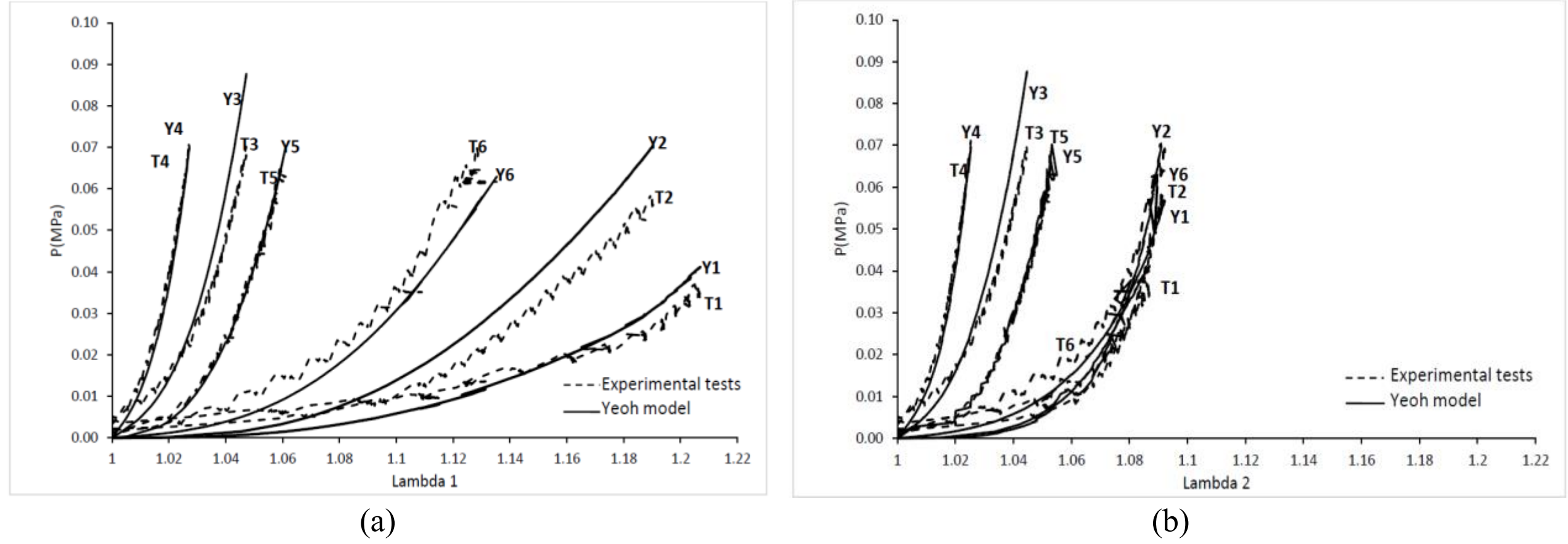

(a) (b)

Fig. 10 - P – stretch ratio curves for experimental data and Yeoh model of the six samples extracted from the left half of the diaphragm. (a) meridional stretch ratio. (b) circumferential stretch ratio. T: test, Y: Yeoh model.

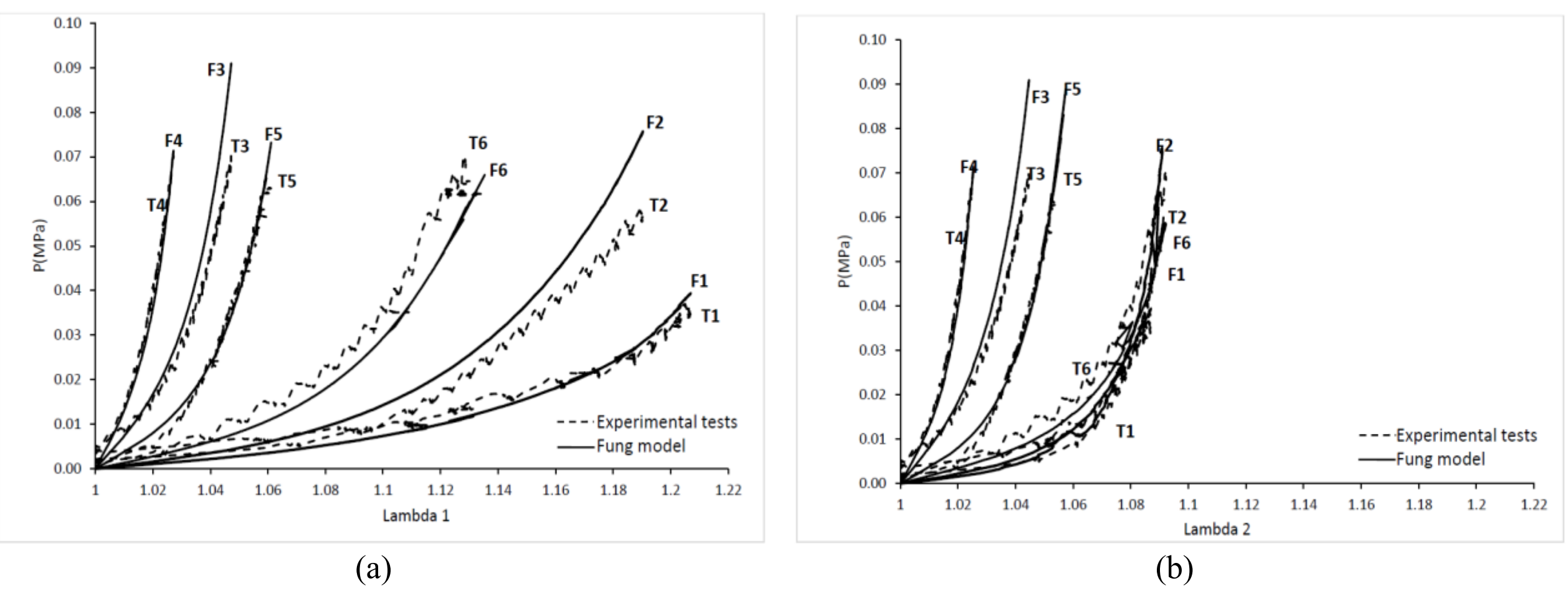

(a) (b)

Fig. 11 - P – stretch ratio curves for experimental data and Fung model of the six samples extracted from the left half of the diaphragm. (a) meridional stretch ratio. (b) circumferential stretch ratio. T: test, F: Fung model.

### *5.2 Transversely isotropic hyperelastic model (Humphrey–Yin)*

In our current study, we initially adopted isotropic models to gain insight into tissue behavior. However, Fung's model emerges as a suitable choice at this stage due to its design for capturing the response of soft tissues rich in collagen fibers. Nonetheless, to better replicate the authentic behavior of the central tendon, it was imperative to integrate models that incorporate the inherent anisotropy of fibrous tissues into our Python code. Noteworthy models, such as the transversely isotropic hyperelastic models developed by Humphrey and Yins [47], and Martins [51], have been developed to address the influence of fibers. On another front, Holzapfel et al [52] formulated a constitutive law to examine the anisotropic and nonlinear elastic behavior of artery walls by considering fiber orientation in both media and adventitia layers. Each layer was treated as a fiber-reinforced material, resulting in an orthotropic constitutive law. To accurately represent anisotropic behavior while accounting for tissue compressibility, Nolan et al [53] proposed a modified version of the previously mentioned model, predicting an anisotropic response under hydrostatic tensile loading, causing a sphere to deform into an ellipsoid. More recently, Chanda et al [54] developed a hyperelastic model that factors in the individual contributions of fiber layers and the matrix, fiber-matrix interaction, and the influence of different fiber orientations within layers.

Although the Humphrey-Yin formulation [47] is often interpreted in terms of explicit fiber reinforcement, it can equivalently be viewed as an invariant-based transversely isotropic hyperelastic model. In the present work, this interpretation is adopted to describe a membrane that is isotropic in-plane but exhibits distinct mechanical behavior through its thickness. In this study, we implemented a transversely isotropic Humphrey-Yin (HY) model, which incorporates fiber stretch via the invariant $I_4$ and provides a mechanistic link between microstructure and macroscopic behavior. Figures 12 and 13 present the experimental pressure–stretch responses together with the predictions of the Humphrey–Yin (HY) model. Compared with all isotropic formulations, the agreement between model and experiment is markedly improved over the entire inflation range.

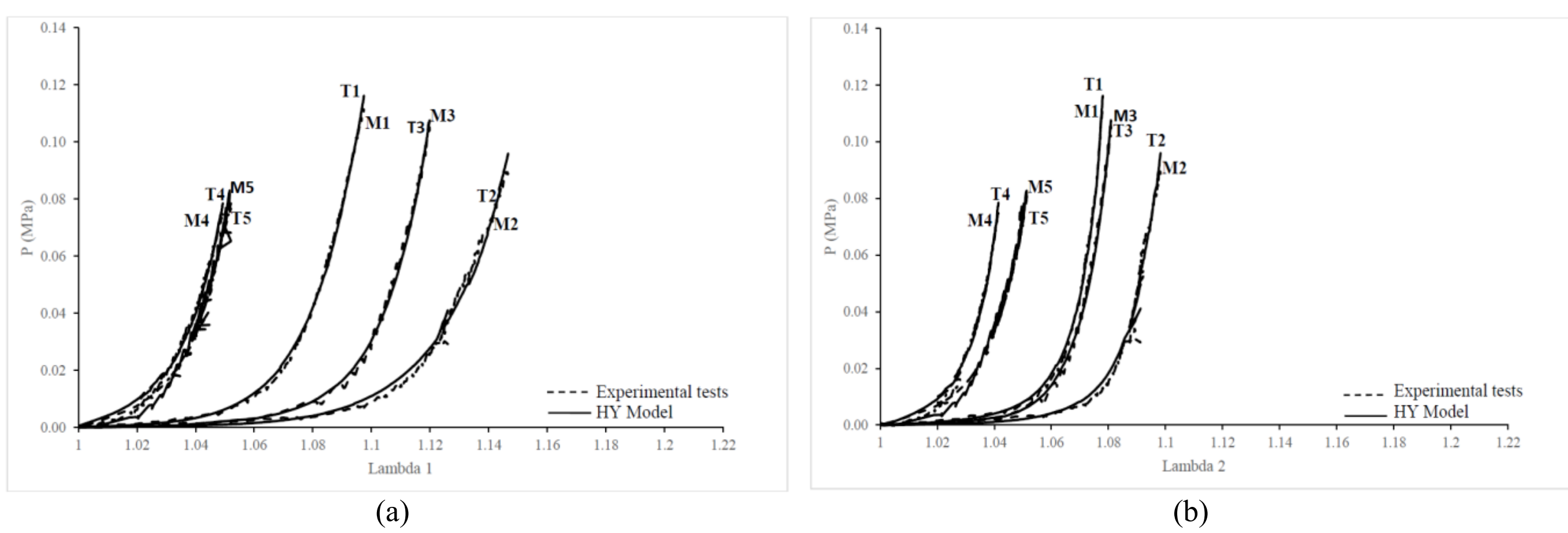

(a) (b)

Fig. 12 - P – stretch ratio curves for experimental data and Humphrey-Yin model of the six samples extracted from the right half of the diaphragm. (a) meridional stretch ratio. (b) circumferential stretch ratio. T: test, M: HY model.

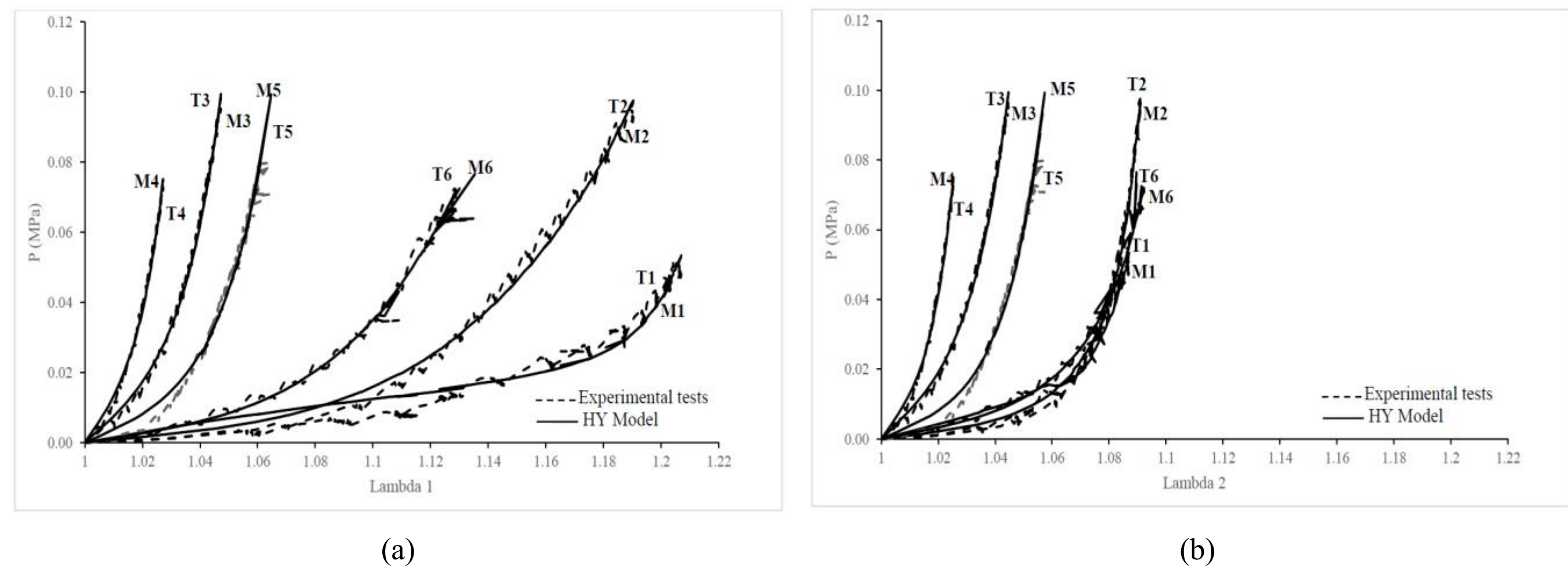

(a) (b)

Fig. 13 - P – stretch ratio curves for experimental data and Humphrey-Yin model of the six samples extracted from the left half of the diaphragm. (a) meridional stretch ratio. (b) circumferential stretch ratio. T: test, M: HY model.

The HY model accurately reproduces the pronounced nonlinear stiffening characteristic of the diaphragmatic central tendon under biaxial inflation. This improvement is observed for both meridional and circumferential responses, despite the apparent axisymmetry of the inflated configuration at the pole. In specimens exhibiting a relatively homogeneous lamellar organization, the Humphrey-Yin formulation provides excellent agreement with the experimental response across all stretch levels. In specimens displaying greater structural heterogeneity, characterized by variations in the number, thickness, and collagen density of lamellar planes, small deviations are observed. These deviations reflect local variability in the stratified architecture of the tendon rather than limitations of the constitutive framework. Importantly, the anisotropic contribution captured by the Humphrey-Yin model should be interpreted as an effective through-thickness response arising from the layered tendon structure. By construction, the in-plane mechanical behavior remains isotropic, and no explicit in-plane fiber reinforcement is introduced. Overall, the Humphrey-Yin model captures both the magnitude of the nonlinear stiffening and the effective anisotropic response associated with the stratified through-thickness organization of the diaphragmatic central tendon under biaxial inflation.

### *5.3 Quantitative comparison: RMSD analysis*

The error between experimental results and Yeoh and Fung hyperelastic models was evaluated using the standard deviation of residuals or Root Mean Square Deviation (RMSD) as follows:

$$RMSD=\sqrt{\frac{\sum residuals^2}{n}} \qquad residuals=\sigma_{exp}-\sigma_{model}$$

Fig. 14 - RMSD between the experimental stress values and the values computed using Yeoh, Fung and Humphrey-Yin (HY) models for all the samples extracted from both halves of the diaphragm.

The results of the RMSD for the eleven samples extracted from both halves of the diaphragm for the two selected models are shown in Fig. (14). Across all specimens, the Fung model outperforms the Yeoh model, yet the HY model consistently yields the lowest RMSD values. These quantitative results are fully consistent with the qualitative comparisons presented earlier and further demonstrate the superior descriptive capability of the Humphrey-Yin formulation for the biaxial inflation response of the diaphragmatic central tendon. Finally, it must be emphasized that the identification of the material parameters within the low stretch ratio domain, as conducted in the current study, often yields unreliable results. This unreliability might stem from inadequate activation of the material's parameters within this domain.

## 6. Conclusions, and Perspectives

This study investigated the biaxial mechanical behavior of the porcine diaphragmatic central tendon using bulge inflation tests combined with full-field stereo digital image correlation. The experimental results revealed a strongly nonlinear hyperelastic response together with pronounced specimen-to-specimen variability, consistent with the heterogeneous and layered organization of the tissue. Although the inflation tests produced a macroscopically axisymmetric, spherical deformation at the pole, classical incompressible isotropic hyperelastic models were unable to consistently reproduce the experimental pressure–stretch relationships. While the Fung model provided improved phenomenological fits over limited stretch ranges, it failed to capture the directional differences observed between meridional and circumferential responses. In contrast, a transversely isotropic Humphrey–Yin formulation, interpreted here as an invariant-based model preserving isotropy within the membrane plane while accounting for through-thickness anisotropy, yielded a markedly improved description of the experimental response over the full inflation range.

The identified material parameters indicate that the anisotropic contribution associated with the stratified tendon structure dominates the strain–energy response. These findings highlight the limitations of isotropic constitutive laws for modeling diaphragmatic tendon mechanics and demonstrate that geometric symmetry of the deformed configuration does not imply material isotropy. More generally, the present work underscores the importance of combining multiaxial experimental loading with full-field kinematic measurements for the identification of constitutive behavior in thin biological membranes. The proposed experimental identification framework provides a robust basis for capturing effective anisotropic responses arising from layered tissue architectures, even when the in-plane deformation appears macroscopically symmetric.

Despite these promising results, the study has several limitations. The assumption of uniform thickness across all samples, without post-test measurements, prevented validation of the incompressibility hypothesis. However, within the applied pressure range, this assumption remains reasonable. The low stretch ratios observed ($\lambda_1$ and $\lambda_2$) suggest that thickness variations may have a limited impact on the results. Another limitation is the inability to assess failure stresses, as tests were terminated once the speckle pattern could no longer be tracked. Additionally, although optical coherence tomography (OCT) imaging was performed, it did not provide sufficient resolution to reveal the collagen fiber architecture. As the samples were fully consumed during testing, no post-experimental structural analysis could be conducted.

Future investigations could build on this approach by integrating complementary microstructural observations, exploring alternative anisotropic constitutive formulations, or extending the analysis to different physiological loading conditions. Such developments would further strengthen the understanding of structure–mechanics relationships in collagen-rich membrane tissues and support the design of biomimetic materials for diaphragmatic repair.

**Conflict of interest**

There is no conflict of interest.

**Acknowledgment**

The authors gratefully acknowledge MUSE International Mobility Program EXPLORE of Montpellier University for funding the stay of Rania ABDELRAHMAN at Montpellier to perform the experimental work.

This work was supported by CNRS - AAP “Osez l’Interdisciplinarité 2018”. MoTiV Project. The authors also thank Gille Camp, Jonathan Bares, Patrice Vallorge and Vincent Huon for their help in the experiments as well as the Association de Chasse Communale Agréée de Gignac - Diane for supplying the wild pig diaphragm.

# Appendix A — DIC Uncertainty Quantification

This appendix provides the detailed procedures and analyses performed to quantify the uncertainty associated with the 3D-DIC measurements used for stretch computation and constitutive parameter identification. Only the main conclusions are presented in the Methods section; the full methodology is documented here for completeness.

## A.1 Speckle Pattern Quality Assessment

For each specimen, the undeformed speckle image was evaluated using a 2D Fourier transform to examine the spatial-frequency content of the stochastic pattern. Following the recommendations of Dong and Pan (2017) [55] , radial spectra were extracted and inspected for:

- absence of periodic low-frequency components,
- isotropy of the frequency distribution,
- sufficient high-frequency content to ensure sub-pixel correlation accuracy.

All specimens exhibited broad and isotropic spectra, consistent with high-quality speckle patterns suitable for DIC on hydrated soft tissues. No sample required exclusion or repainting.

## A.2 Illumination and Imaging Stability

Because hydrated biological tissues are prone to specular reflections and contrast drift, illumination conditions were controlled throughout the test campaign:

- backlighting intensity was kept constant,
- camera exposure and gain were fixed for all acquisitions,
- grayscale histograms were monitored to detect saturation or contrast loss.

No significant variation in histogram distribution or dynamic range was observed across tests. In particular, no degradation of contrast occurred during the inflation steps, confirming the reproducibility of the optical conditions.

## A.3 Stereo Reconstruction Accuracy and Calibration Quality

The stereo system was calibrated before each testing session using a dot-grid target. Calibration residuals remained below 0.03 pixels for all sessions. This value is consistent with high-quality stereo-DIC configurations and below typical thresholds affecting strain estimation.

Three additional indicators were monitored:

1. Reprojection error: stable across calibrations.
2. Epipolar consistency: no measurable drift across trials.
3. Rigid-body checks on undeformed images: no spurious displacement fields observed.

These metrics confirm the stability of the camera geometry and reconstruction mapping.

## A.4 Numerical Sensitivity Analysis Using Synthetic Data

To quantify the influence of geometric measurement noise on constitutive parameter identification, synthetic bulge-test datasets were generated from analytical membrane solutions. The following procedure was applied:

1. A reference spherical deformation was computed for a Neo–Hookean membrane with known material parameter.
2. Random perturbations of up to ±20% were added to the out-of-plane displacement field to mimic DIC-related noise.
3. Perturbed surfaces were reprocessed to extract curvature, stretches and pressure–stretch curves.
4. The identification algorithm was applied to recover the material parameter.

Across all perturbation levels, the relative deviation of the identified parameter remained below **2%**, demonstrating strong robustness of the stretch computation and identification pipeline to geometric noise.

## A.6 Summary of DIC-Related Uncertainties

Considering all specimens and analyses conducted, the following conclusions can be formulated regarding the uncertainty associated with the 3D-DIC measurements used in this study:

- The speckle patterns were visually uniform and exhibited frequency characteristics compatible with reliable correlation.
- Illumination and imaging conditions were maintained constant throughout the experiments, and no systematic contrast drift was observed.
- Stereo calibration residuals remained low and stable across testing sessions, indicating a reliable reconstruction geometry.
- No systematic artifacts attributable to decorrelation were identified during image processing.
- Numerical sensitivity analyses based on synthetic data indicate that moderate geometric perturbations induce only limited variations in the identified constitutive parameters.

Taken together, these observations suggest that measurement-related uncertainties associated with the 3D-DIC procedure are sufficiently small to support the reported stretch measurements and parameter identification, and that the observed inter-sample variability predominantly reflects intrinsic biological heterogeneity rather than experimental noise.